\documentclass[namedreferences]{kluwer}

\usepackage{graphicx}

\newdisplay{guess}{Conjecture}

\begin{document}

\begin{article}
\begin{opening}
\title{Scaling Relations of Field Spirals \\ 
at intermediate Redshift\thanks{Based on observations with the 
ESO Very Large Telescope,
run IDs 65.O-0049, 66.A-0547 and 68.A-0013.}}
\author{A. \surname{B\"ohm}\email{boehm@uni-sw.gwdg.de}}
\author{B.L. \surname{Ziegler}}
\author{K.J. \surname{Fricke}}
\institute{Universit\"atssternwarte G\"ottingen, Geismarlandstr. 11,
37083 G\"ottingen, Germany}
\author{and  \surname{the FDF Team}}
\institute{Landessternwarte Heidelberg, Universit\"atssternwarte M\"unchen}

\runningauthor{A. B\"ohm, B.L. Ziegler, K.J. Fricke and the FDF Team}
\runningtitle{Scaling Relations of Field Spirals at intermediate Redshift}

\begin{abstract}
In the last few years, galaxies at redshifts up to $z \sim 1$ have become
accessible for medium--resolved spectroscopy thanks to the new generation
of 10m-class telescopes. With kinematic and photometric information on 
spiral galaxies in this regime, well--known scaling relations like the 
Tully--Fisher relation (TFR) can be studied over half a Hubble time. 
By comparison to local samples, these studies facilitate simultaneous tests
of  the hierarchical merging scenario and stellar population models. \\
Using the Very Large Telescope, we obtained spatially resolved rotation 
curves of 78 spiral galaxies in the FORS Deep Field (FDF), covering all Hubble 
types from Sa to Sm/Irr at redshifts $0.1 < z < 1.0$. 
We find evidence for a $B$-band luminosity increase of up to 2\,mag
for low--mass spirals, whereas the most massive galaxies are of the 
same luminosity as their local counterparts. In effect, the 
TFR slope decreases significantly. This would explain the discrepant
results of previous observational studies. 
We also present the velocity--size relation and compare it to
the predictions of numerical simulations based on the 
hierarchical merging scenario.

\end{abstract}

\keywords{galaxy evolution, cosmology}

\end{opening}

\section{Motivation and Sample Selection}

The scaling relations between the basic parameters of spiral galaxies
--- luminosity $L$, maximum rotation velocity $V_{\rm max}$ and scalelength
$r_{\rm d}$ --- are correlated via a two--dimensional plane which is
similiar to the well-known Fundamental Plane for ellipticals 
(e.g., \opencite{KSW2000}). Most famous
among the projections of this plane is the Tully--Fisher relation (TFR)
between $V_{\rm max}$ and $L$ \cite{TF1977}.
A study of such scaling relations over different cosmic epochs offers
powerful tests of different aspects of the hierarchical merging scenario
and stellar population models.

Nevertheless, spectroscopy of galaxies at redshifts up to  $z=1$ with
sufficient (spectral and spatial) resolution and S/N for gaining robust
information on their kinematics is a great observational challenge and has
become feasible just within the last few years with 10m-class telescopes.
Additionally, any selection on emission line strength or disk size is 
likely to introduce biases in the results. 
%
%
To avoid this, we selected our targets purely on apparent magnitude $R<23^m$
and inclination $i>40^\circ$. Our source was the FDF photometric redshifts 
catalogue (\opencite{A2000}, \opencite{B2001}). 

Throughout this article, we assume the concordance cosmology with
$\Omega_{\rm m} = 0.3$, $\Omega_\Lambda = 0.7$ and
$H_0 = 65$\,km\,s$^{-1}$\,Mpc$^{-1}$.

\section{Analysis}

We derived rotation curves (RCs) by applying Gaussian fits row by row to the
usable emission lines in our spectra. After the rejection of disturbed or 
``solid-body'' RCs, the final sample consisted of 78 spirals at a median
redshift of 0.45. 

To derive proper intrinsic rotation velocities, we performed 
simulations of the spectroscopy for each galaxy by generating synthetic
velocity fields. With these, we corrected for observational effects
like  seeing, disk inclination, angle between slit and
major axis, and also the optical ``beam smearing'' which originates
from the comparable sizes of the slit width (one arcsecond) and the
galaxies' apparent radii. Absolute $B$-band magnitudes were derived via
synthetic photometry
from observed $B$, $g$, $R$ or $I$, depending on the redshift, to keep 
the k-corrections small. For a more detailed description and a data table
of our spiral sample, see \inlinecite{B2002}.  

\section{Results}

The final $B$-band TFR is shown in figure 1. Only RCs which show a
region of constant rotation velocity (due to the Dark Matter Halo) were used
for the bootstrap fit to our sample. In comparison to the local sample of
\inlinecite{Haynes}, which comprises $\sim$\,1200 spirals, 
we find 2$\sigma$$+$ evidence for a
change of the TFR slope at intermediate redshift. This may be caused by a
mass--dependent luminosity evolution which is stronger for lower mass
systems, possibly combined with 
an additional population of blue, low--mass galaxies which is
underrepresented in the local universe. As early--type spirals have, on the
mean, higher maximum rotation velocities than late--type spirals, our result
offers an explanation for the discrepant results of earlier studies, which
were limited to small number statistics and mostly biased towards certain
sub--types like low--mass systems with blue colors (e.g., \opencite{Rix})
or, in contrast, luminous early--type spirals due to the selection upon
large scalelengths \cite{Vogt96}.

In figure 2 we show the velocity--size diagram of our sample.
The slightly decreased disk sizes of the FDF spirals are in good agreement
with the predictions of the hierarchical merging scenario following 
\inlinecite{MMW98}, though the scatter is relatively large since
the imaging is ground--based yet. 
We will improve the accuracy of the measured scalelengths and inclinations 
with HST-ACS observations during cycle 11.

%
%
\begin{figure}[t]
\vspace*{-0.0cm}
\hspace*{-0.3cm}
\centerline{\includegraphics[width=10.5cm]{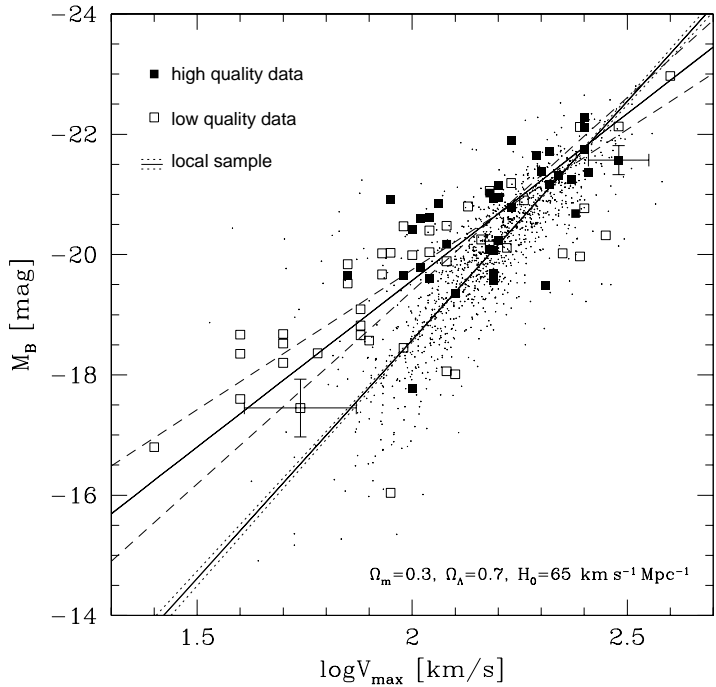}}
\caption{The $B$-band Tully--Fisher relation of our intermediate redshift
sample compared to the $N=1200$ local sample of Haynes et al. (1999). 
Typical error bars are shown for two objects.
Solid lines show the 100 iteration bootstrap bisector fits along with 
1$\sigma$ errors (dashed and dotted lines). 
Only rotation curves which show a constant rotation
velocity at large radii are used for the fit to the FDF data. 
Both samples are corrected
for incompleteness bias and morphological bias following Giovanelli et al.
(1997). A TFR slope change between the local universe and intermediate
redshift is found on $>$2$\sigma$ confidence level.}
\end{figure}

%
%
\begin{figure}[t]
\vspace*{-0.0cm}
\hspace*{-0.3cm}
\centerline{\includegraphics[width=10.5cm]{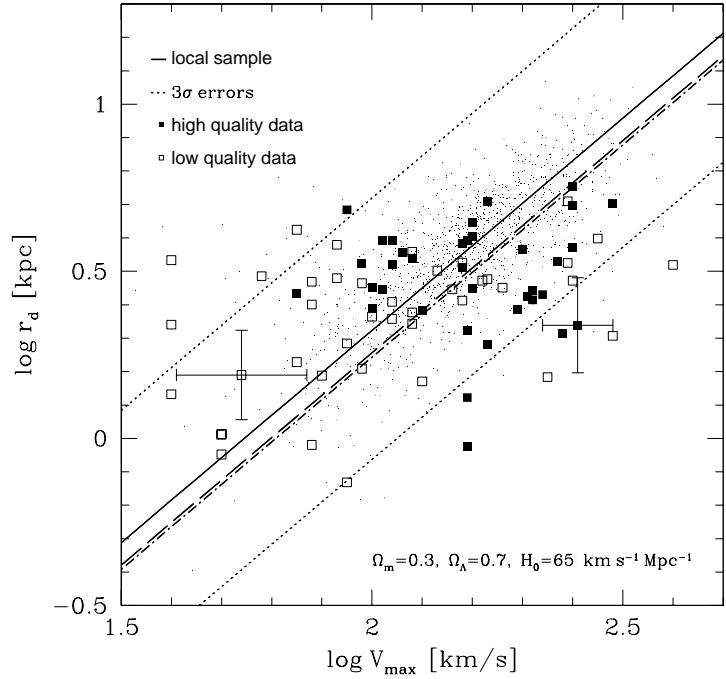}}
\caption{
The velocity--size diagram for our intermediate redshift sample compared to 
the $N=1200$ local sample of Haynes et al. (1999). 
Typical error bars are shown for two objects.
The solid and long--dashed lines give the bisector fits to the local and
intermediate redshift samples, respectively. 
The dot--dashed line denotes the predicted smaller disk sizes 
at $z \sim 0.5$ within the $\Lambda$CDM  hierarchical merging model
following Mao, Mo and White (1998).
}
\end{figure}

\acknowledgements
We thank ESO and the Paranal staff for the efficient support of the 
observations.
This work was funded by the Volkswagen Foundation.

\end{article}
\end{document}